\documentclass[aps,twocolumn,pra,tightenlines,floatfix,showpacs]{revtex4}
\usepackage[dvips]{graphicx}
\usepackage[english]{babel}
\usepackage{amsmath}
\usepackage{amssymb}
\usepackage{times}

\newcommand{\bs}{\begin{split}}
\newcommand{\es}{\end{split}}
\newcommand{\be}{\begin{equation}}
\newcommand{\ee}{\end{equation}}
\newcommand{\ba}{\begin{eqnarray}}
\newcommand{\ea}{\end{eqnarray}}

\def\ket#1{|#1\rangle}

\begin{document}

\title{Radio Frequency Spectroscopy of Trapped Fermi
Gases with Population Imbalance}

\author{Yan He, Chih-Chun Chien, Qijin Chen, and K. Levin}

\affiliation{James Franck Institute and Department of Physics,
 University of Chicago, Chicago, Illinois 60637}

\date{\today}

\begin{abstract}

  Motivated by recent experiments, we address, in a fully self
  consistent fashion, the behavior and evolution of radio frequency (RF)
  spectra as temperature and polarization are varied in population
  imbalanced Fermi gases. We discuss a series of scenarios for the
  experimentally observed zero temperature pseudogap phase and show how
  present and future RF experiments may help in its elucidation.  We
  conclude that the MIT experiments at the lowest $T$ may well reflect
  ground state properties, but take issue with their claim that the
  pairing gap survives up to temperatures of the order of the degeneracy
  temperature $T_F$ at unitarity.

\end{abstract}
\pacs{03.75.Hh, 03.75.Ss, 74.20.-z \hfill \textbf{\textsf{arXiv:0707.2625}}}

\maketitle

The field of ultra-cold Fermi gases undergoing BCS-BEC crossover is
particularly exciting because these superfluids exhibit a rather novel
form of fermionic superfluidity: pairing begins at temperature $T^*$
while condensation take place at a significantly lower temperature
$T_c$.  Concomitantly, the normal state fermionic spectrum exhibits an
excitation gap or ``pseudogap'' \cite{Grimm4,Reviews,ThermoScience}.
Experiments from MIT \cite{ZSSK06} on population imbalanced Fermi gases
simultaneously observe vortices and condensate fractions to estimate
$T_c$. Combined with RF spectroscopy \cite{KetterleRF} they have thereby
established that a pairing gap is indeed visible \textit{above} $T_c$.
An even more striking claim from this group \cite{KetterleRF} is that in
a highly polarized gas, one finds a ground state in which $T^* \neq 0$
while $T_c =0$.  Throughout this paper we refer to this state as ``a
zero temperature pseudogap phase''.

It is the goal of the present paper to address these RF spectroscopy
experiments on spin imbalanced \cite{Rice1,Rice2,MITPRL06,ZSSK06}
unitary Fermi gases.  Our aim is to show theoretically how the self
consistently calculated RF spectra evolve as temperature and
polarization are varied thereby accessing the various phases which have
been contemplated \cite{Rice2,ChienPRL,Parish}. With (roughly)
decreasing $T$, these correspond to a Fermi gas, a pseudogap phase, a
polarized superfluid (or ``Sarma'' state) and a phase separated state.
The latter is the ground state for all but possibly the highest
polarizations at unitarity \cite{MITPRL06,Rice1}.  An additional aim is
to discuss a series of (four) scenarios for the important zero
temperature pseudogap phase and show how present and future RF
experiments may help to clarify what is going on.  In examining one of
these we conclude that the lowest $T$ in the MIT experiments may, as
argued, reflect ground state properties.  Nevertheless we take issue
with another claim from this group-- that the pairing gap survives up to
temperatures of the order of the degeneracy temperature $T_F$.  This
high value for $T^*$ at unitarity appears inconsistent with all existing
crossover theories, and with earlier thermodynamical indications
\cite{ThermoScience}.

Because it includes the pseudogap in a self-consistent fashion, we use a
theoretical approach \cite{Reviews,Chen4,JS2} to BCS-BEC crossover which
appears to be uniquely positioned to address RF experiments
\cite{Torma2,heyan} on polarized gases \cite{Chien06,ChienPRL} at
general temperatures $T$.  Presuming, as we do here, the standard
BCS-Leggett ground state, then Bogoliubov-de Gennes (BdG) based schemes
are also applicable \cite{BdG_Imbalance} but only to strictly $T=0$,
where the non-condensed pairs of central interest here, do not enter.
Essentially all other schemes in the literature, inspired by the
Nozi\`eres--Schmitt-Rink approach \cite{NSR}, contain a problematic
inconsistency in their incorporation of these pseudogap effects. They
presume that [in the fermionic dispersion relation $E_{\bf k} = \sqrt{
  \xi_{\bf k}^2 + \Delta^2 (T) }$] the pairing gap $\Delta$ vanishes at
and above $T_c$.

In this paper we use the standard one channel grand canonical
Hamiltonian $H-\mu_1 N_1-\mu_2 N_2$ which describes pairing between
states $\ket1$ and $\ket2$ and for definiteness take state $\ket1$ as
majority and state $\ket2$ as minority, unless indicated otherwise.
We additionally ignore the interaction between state $\ket3$ and states
$\ket1$ and $\ket2$, since mean-field energy shifts associated with the
interaction between $\ket1$ and $\ket2$ and between $\ket1$ and $\ket3$
nearly cancel each other, as observed experimentally. Thus state $\ket3$
is associated with a noninteracting gas.
In addition, there is a transfer matrix element $T_{\bf k,p}$ from
$\ket2$ to $\ket3$ given by
$  H_T  =\sum_{\bf k,p}(T_{\bf k,p}\,c_{3,\bf p}^\dag c_{2,\bf
k}^{} +h.c.) .$
For plane wave states,
$T_{\bf k,p} = \bar {T }\delta({\bf q}_L+{\bf k}-{\bf
p})\delta(\omega_{\bf{kp}} - \omega_L)$.
Here $q_L \approx 0$ and $\omega_L$ are the momentum and energy of the
RF laser field, and $\omega_{\bf {kp}}$ is the energy difference between
the initial and final state.  It should be stressed that unlike
conventional quasi-particle tunneling, here one requires not only
conservation of energy but also conservation of momentum.

The RF current is defined as $I = \langle \dot{N}_3\rangle=i\langle
[H,N_3]\rangle $.  Using standard linear response theory 
one finds
\begin{eqnarray}
  I&=& 2 \bar{T}^2{\rm
    Im}[X_{ret}(-\omega_L+\mu_3-\mu_2)] ,\nonumber \\
  X(i\omega_n) &=& T
  \sum_{m,{\bf k}}G_3({\bf k},i\nu_m)G_{2}({\bf
    k+q}_L,i\nu_m+i\omega_n)\,,
\end{eqnarray}
where $\mu_3$ is the chemical potential of $\ket3$ and $\omega_{23}$ is
the energy splitting between $\ket3$ and $\ket{2}$.
After Matsubara summation and using $A_3({\bf k},\nu)=2\pi
\delta(\nu-(\epsilon_{\bf k}+\omega_{23}-\mu_3))$ as well as $A_{2}({\bf
  k},\nu) \equiv -2\, \mathrm{Im}\, G_{2}({\bf k},\nu+i 0^+)$ to rewrite
the spectral functions for states $\ket3$ and $\ket2$, respectively, we
have
\ba
 I(\omega)&=&\frac{\bar{T}^2}{2\pi}\sum_{\bf k}
A_{2}({\bf k}+{\bf q}_L,\epsilon_{\bf k}-\omega-\mu_2)\nonumber\\
&&{} \times\left[f(\epsilon_{\bf k}-\omega-\mu)-f(\epsilon_{\bf
k}+\omega_{23}-\mu_3)\right] \,, \label{eq:4} \ea
where $\omega\equiv \omega_L - \omega_{23}$ is defined to be the RF
detuning and $f(x)$ is the Fermi distribution function.  In the above
equations the retarded response function $X_{ret}(\omega) = X(i\omega_n
\rightarrow \omega+i0^+)$, and we have expressed the linear response
kernel $X$ in terms of single particle Green's functions.  We define
$\omega_n$ and $\nu_m$ as even and odd Matsubara frequencies,
respectively and $G_{2}$ is the fully dressed Greens function for the
state $2$ spins. (We use the convention $\hbar=k_B=1$).

In our $T$-matrix formalism \cite{Chen4,JS2}, 
$G_{2}(\mathbf{k},\nu)$ contains two self-energy contributions deriving
from condensed Cooper pairs ($\Sigma_{sc}$) as well as from finite
momentum pairs ($\Sigma_{pg}$). The latter represent pseudogap effects
which first appeared in the spectral function, in Ref.~\cite{Janko}.
We have $\Sigma = \Sigma_{pg} + \Sigma_{sc}$, where 
$\Sigma_{pg} ({\bf k}, \nu) = \frac{\Delta_{pg}^2}{\nu
  +\xi_{k,1} +i\gamma} $ and $\Sigma_{sc}({\bf k}, \nu) =
\frac{\Delta_{sc}^2}{\nu+\xi_{k,1}}$.
  Here $\Delta_{sc}$ is the superfluid order parameter, and $\gamma \ne
  0$ is associated with the life time effects of noncondensed pairs. The
  resulting spectral function can readily be computed as
\be A_2({\bf k},\nu)=\frac{2\Delta_{pg}^{2}\gamma
  (\bar\nu+\xi_{\bf k})^2}{(\bar\nu+\xi_{\bf
    k})^2(\bar\nu^2-E_{\bf k}^{2})^2+\gamma^2(\bar\nu^2-\xi_{\bf
    k}^2-\Delta_{sc}^{2})^2} \,.
\label{eq:5} \ee
Here $\xi_{{\bf k},1} = \epsilon_{\bf k}-\mu_1$, $\xi_{\bf k} =
\epsilon_{\bf k}-\mu$, $\mu = (\mu_1+\mu_2)/2$, $h=(\mu_1-\mu_2)/2$, and
$\bar\nu=\nu-h$.  In the quasiparticle dispersion, $E_{\bf k}$,
$\Delta^2(T) = \Delta_{sc}^2(T) + \Delta_{pg}^2(T)$.  The precise value
of $\gamma$, and even its $T$-dependence is not particularly important,
as long as it is non-zero at finite $T$. In practice, we choose its
value based on the experimental atomic peak width. As is consistent with
the standard ground state constraints, $\Delta_{pg}$ vanishes at $T
\equiv 0 $, where all pairs are condensed.  Above $T_c$, we have
Eq.~(\ref{eq:5}) with $\Delta_{sc} = 0$.  Because the energy level
difference $\omega_{23}$ ($ \approx 80 $ MHz) is so large compared to
other energy scales in the problem, the state $\ket3$ is initially empty
and thus $f(\epsilon_{\bf k}+\omega_{23}- \mu_3) = 0$ in
Eq.~(\ref{eq:4}).  Once the trap is incorporated, Eqs.~(\ref{eq:4}) and
(\ref{eq:5}) can then be used to compute the local current density $I(r,
\omega)$ and then to obtain the total net current
%\begin{equation}
$I_{\sigma}(\omega) = \int \mathrm{d}^3r\, I(r, \omega)n_{\sigma}$
%\end{equation}
with $\sigma=1,2$.  Unless stated otherwise the energy unit $T_F$
represents the Fermi temperature for the noninteracting unpolarized
Fermi gas with the same total particle number.

To treat the trap, we assume a spherically symmetrical harmonic
oscillator potential $V(r) = m\bar{\omega}^2 r^2/2$.  The density,
excitation gap and chemical potential will vary along the radius. These
quantities can be self-consistently determined using the local density
approximation (LDA).  The phase diagram, representing the stable regimes
for phase separation, the Sarma phase as well as the normal Fermi gas
phases as a function of temperature and polarization has been mapped out
\cite{Parish,Rice2}. Since it is at the heart of the current
experments, one must also determine \cite{ChienPRL} where
\textit{pairing occurs without superfluidity}.  These non-condensed pair
effects (which are generally ignored in the literature) are also
essential for arriving at physical values for $T_c$.  Important for the
present purposes, the phase separated state is \textit{not} associated
with pseudogap effects, unlike the Sarma state.  The same behavior is
mirrored in the density profiles. The Sarma phase consists of a
superfluid core followed by a correlated ``mixed normal'' or pseudogap
regime, followed by a Fermi gas in the outer regions of the trap. The
phase separated state, by contrast has an essentially unpolarized
superfluid core separated from a noncorrelated normal Fermi gas by a
sharp interface.  The phase boundary is determined \cite{ChienPRL} by
the balance of pressure.

\begin{figure}
\centerline{\includegraphics[clip,width=2.62in]{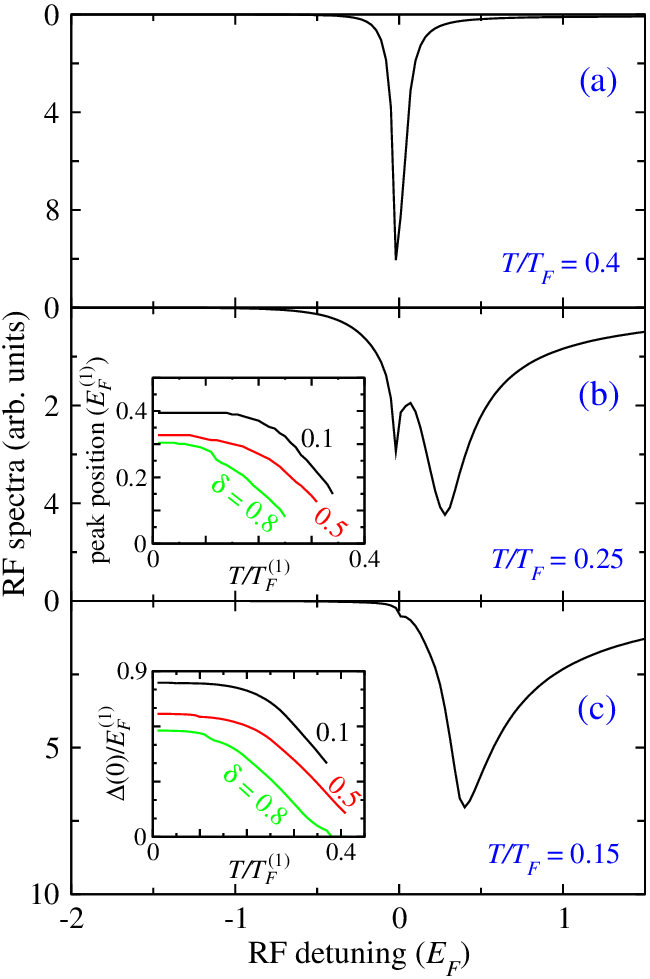}}
%\centerline{\includegraphics[clip,width=2.62in]{rfa0p0.5new.eps}}
\caption{RF spectra for polarized gases in a harmonic trap at unitarity
  and polarization $\delta = 0.5$, for (a) $T/T_F = 0.4$, (b) 0.25, and
  (c) 0.15, respectively. The insets in (b) and (c) are, respectively,
  the pairing peak position and the energy gap $\Delta(T)$ at the trap
  center as a function of $T/T_F^{(1)}$ (in units of the majority Fermi
  energy $E_F^{(1)}$), for $\delta=0.1$ (black), $0.5$ (red), and $0.8$
  (green lines), as labeled. The corresponding $T_c/T_F = 0.28$ (black),
  $0.25$ (red), and $0.19$ (green lines), respectively, and the
  estimated $T^*$ can also be read off from the insets where the gap
  vanishes. Here we choose $\gamma=0.05$.}
\label{rfa0p0.5}
%\label{fig:1}
\end{figure}

To begin, it is useful to present the prototypical behavior for the RF
spectra.  Quite generally we find that in the phase separated state (low
$T$) there is a single pairing peak, whereas in the pseudogap phase
(higher $T$) there are two peaks.  And the Sarma phase (intermediate
$T$) may have either one or two, depending on $T$ and $\delta$.
Finally, at high $T$, we have only an atomic peak, located precisely at
the atomic level separation $\omega_{23} =0$.  For a range of lower $T$,
the atomic peak persists deriving from the effectively noninteracting
Fermi gas contribution at the trap edge; the pairing peak arises from
the superfluid or pseudogap region in the trap center.

Figure \ref{rfa0p0.5} presents self consistent numerical results for the
minority RF spectra at unitarity and at moderate polarizations $\delta =
(N_\uparrow -N_\downarrow)/N = 0.5$.  The temperature gradually
increases from the lower to upper panels.  We consider lower
temperatures (by a factor of about $2$) to arrive at results which are
comparable to those in \cite{KetterleRF}.
The two insets show the pairing peak position and trap center gap as a
function of $T/T_F^{(1)}$.  In these two insets, we follow the MIT
experiments and use the majority component Fermi energy $E_F^{(1)}$ as a
unit of energy for both temperature and gap. The black, red and green
curves correspond to three polarizations: $\delta=0.1$, $0.5$ and $0.8$
respectively. One can see that the higher polarization is associated
with a smaller peak position and energy gap.  We see that the magnitudes
of the pairing gap are rather comparable to their experimental
counterparts.  As in experiment the pairing gap increases with
decreasing temperature. The energy scale at which it smoothly vanishes
can be read off in the insets which yield $T^*$.  There is no sharp
feature at $T^*$, so experimentally it cannot be precisely defined.
Nevertheless, we see that there is a clear separation between the peak
location curves for the three polarizations. By contrast the
experimental data for all measured polarizations lie on the same
(approximately) universal curve, with substantially higher $T^*$ (by a
factor of $2$ or so).

\begin{figure}
\centerline{\includegraphics[clip,width=3.3in,height=3.in]{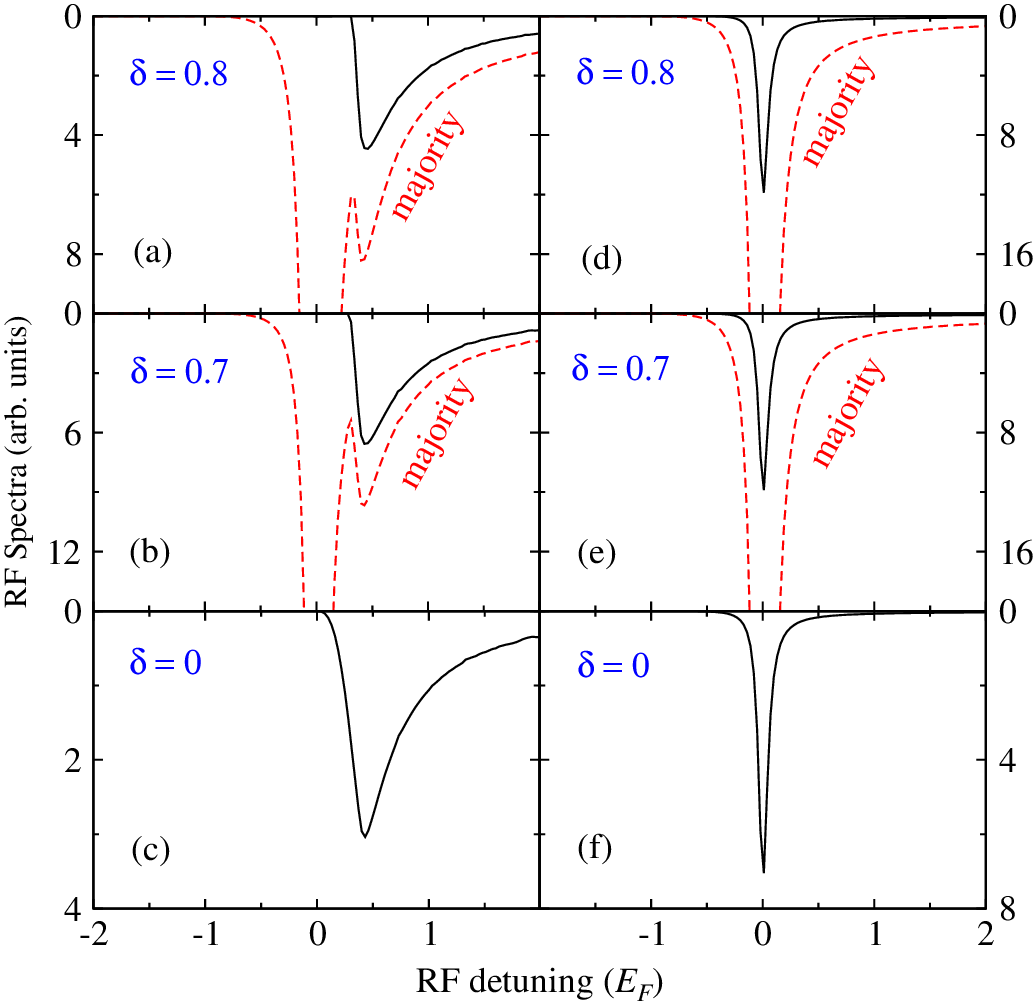}}
%\centerline{\includegraphics[clip,width=3.3in,height=3.in]{rfa0T0.01new.eps}}
\caption{Low temperature ($T=0.01$) RF spectra at different
  polarizations (as labeled) for a unitary (left) and noninteracting
  Fermi gas.  When $\delta \neq 0$, the solid (black) and dashed (red)
  curves show the result when state $\ket2$ is the minority and
  majority, respectively.}
\label{rfa0T0.01}
\end{figure}

We now turn to a first scenario for elucidating the exotic
non-superfluid phase at high polarizations \cite{KetterleRF} by
considering the possibility \cite{Lobo_Chevy2} that this state is a
Fermi gas or liquid.  The loss of superfluidity would be due to a
destabilization (arising from more benign Hartree-like corrections) in
the competing normal Fermi gas phase.  This scenario is not compatible
with a zero temperature pseudogap phase (since the presence of an
excitation gap for fermions means that it is not in a Fermi gas or Fermi
liquid state). Nevertheless, this scenario would give rise to a single,
nearly symmetric RF peak at low temperatures and high polarizations,
similar to that observed experimentally, albeit associated with an
atomic rather than pairing peak.

Figure \ref{rfa0T0.01} plots the self consistently determined RF spectra
at very low temperatures $T=0.01T_F$ in the unitary (left) and the
noninteracting limit (right panels), assuming state $\ket2$ is the
majority (red dashed) and minority (black solid lines), respectively.
The top two panels correspond to high polarizations, $\delta=0.7$ and
$0.8$. The bottom panel presents a comparison with an unpolarized gas.
This low temperature phase corresponds to superfluidity in all cases in
the left column, since that is what is found in our self consistent
calculations.  The two high polarizations correspond to phase
separation.  In the noninteracting gas case (right column), the results
are very simple.  We find, as expected, only atomic peaks in the
majority and minority curves. They are located at precisely the same
position -- at the zero of our frequency scale.
Comparing the two curves in (a) and (b) with (d) and (e) one sees that
with future majority spectra there is a simple way to rule out this
particular Fermi gas scenario.  At low $T$ the majority curves in (a)
and (b) (unlike the minority) have atomic peaks as well as pairing
peaks.  The larger atomic peaks of the majority plots are associated
with the fact that the majority has a much larger noninteracting gas
tail in its particle density profile.  By contrast for the minority
curves on the left, all fermions are paired at these low $T$ and we see
only a single pairing peak.

When comparing with existing experiments, it should be noted that if the
single peak in the zero temperature pseudogap phase were an atomic peak
such as in the calculations of Ref.~\cite{Lobo_Chevy2}, there would be
a shift in its position (relative to that computed here) though probably
not large enough to match the experimental presumed pairing peak.  In
summary, \textit{this figure shows that the combined measurement of both
  majority and minority curves can serve to establish whether a single
  peak is coming from paired atoms or noninteracting atoms}. In this way
it can address the scenario which associates the non-superfluid state at
high polarizations with a Fermi gas phase.

\begin{figure}
\centerline{\includegraphics[clip,width=2.5in]{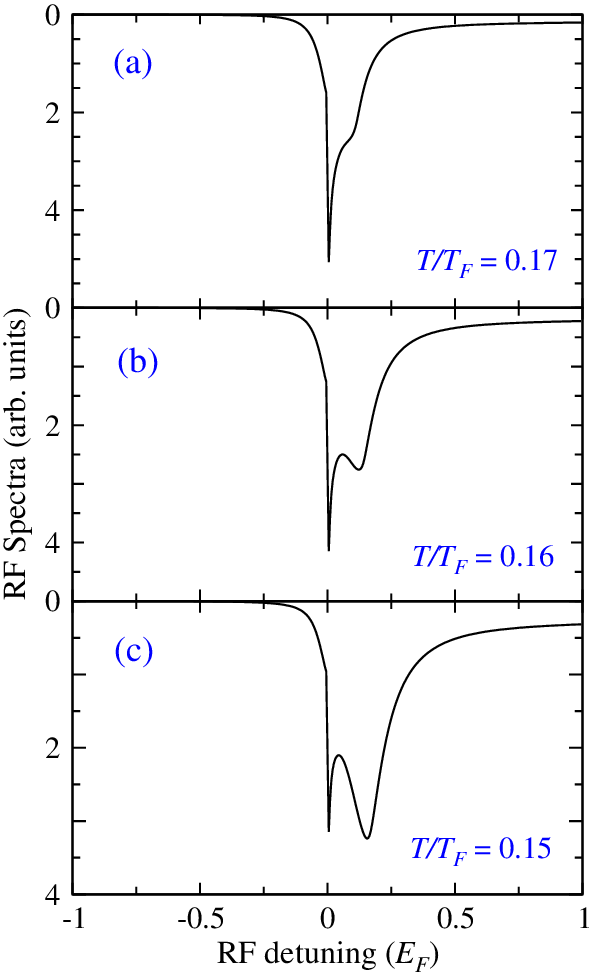}}
%\centerline{\includegraphics[clip,width=2.5in]{rfa0p0.95.eps}}
\caption{RF spectra at unitarity when $\ket2$ is the minority with
  polarization $\delta=0.95$, for different temperatures (a) $
  T/T_F=0.17$, (b) $0.16$, and (c) $0.15$, respectively. Here we take
  $\gamma=0.05$.}
\label{rfa0p0.95}
%\label{fig:3}
\end{figure}

In Fig.~\ref{rfa0p0.95}, we turn to another possible scenario for the
mysterious zero temperature pseudogap phase and calculate the self
consistent RF spectra within a \textit{finite temperature (normal)
  pseudogap phase} which arises in the Sarma portion of the phase
diagram at very high polarizations \cite{ChienPRL}.  Here we consider
$\delta=0.95$, in order to have polarization and temperatures consistent
with the computed phase diagram \cite{ChienPRL}.  The temperature
gradually increases from $T=0.15T_F$ to $0.17T_F$ as we go from bottom
to top panels.  This figure is based on the implicit possibility that
the purported $T=0$ pseudogap phase \cite{KetterleRF} is a finite
temperature observation. We thus use RF experiments as a type of
thermometry and probe whether the experimental temperatures are
sufficiently low to be in the true ground state.  Importantly, we see
from the figure that a two peaked structure is clearly visible at the
lowest $T$ of this intermediate temperature scale, $0.15T_F$.  It will
be even better resolved at somewhat lower polarizations, as studied
experimentally.
The two peaks start to merge at $0.17T_F$, where we are left with an
atomic peak only.  The observation of two peaks in this figure, in
contrast with experiment, \textit{suggests that the MIT
experiments were conducted at sufficiently low $T$}.

A third possible scenario for the observed zero temperature pseudogap
phase follows from BdG-based calculations \cite{BdG_Imbalance} which
suggest that the ground state is not phase separated as in LDA theories
at unitarity, but instead a superfluid with a complex order parameter --
in a Fulde-Ferrell-Larkin-Ovchinnikov (FFLO) state \cite{FFLO}.  We have
conducted a finite temperature study (importantly including noncondensed
pairs) of the simplest such state \cite{LOFF1} which suggests that this
oscillatory order parameter phase rapidly becomes unstable with
increasing temperature. Because it is not sufficiently robust, we argue
that the FFLO phase is not likely to be a candidate for the exotic
ground state.  Indeed, experiments from both groups seem to support
phase separation \cite{MITPRL06,Rice1}.  We stress that the phase
separation that is consistent with the current theory seems to be more
akin to that in Ref.~\cite{Rice1} than that in Ref.~\cite{MITPRL06}
where there is very little, if any, pseudogap regime, in either the
density profiles or the phase diagram.

As a fourth scenario, we note that the most natural way to obtain $T_c=
0$ with $T^* \neq 0$ is associated with phases in which there is a
frustration of pair mobility which leads to localization of pairs.  This
state appears in recent theoretical work on unitary gases
\cite{Chienlattice} in the presence of optical lattices.  Nevertheless,
it appears difficult to understand how it can arise from Zeeman-like
effects, which primarily break pairs apart.  Most likely (but for very
different reasons) this phase has been observed in high $T_c$
superconductors under various perturbations where $T_c$, but not $T^*$,
is driven to zero \cite{Reviews}.

In summary, this paper has shown that future RF experiments are needed
to arrive at a more conclusive understanding of the observed pairing
peaks, hopefully, both by reducing the unexpectedly high estimates of
$T^* \gtrsim T_F$ (in order to be consistent with essentially all
estimates), and via providing majority spectra.  The latter can confirm
the presence of a pairing, as contrasted with, an atomic peak.  We
cannot rule out the possibility that the purported $T=0$ pseudogap phase
has some form of superfluid order. However, if instead a non-superfluid
but paired ground state is confirmed, it will very likely contain some
degree of ``bosonic'' order.

We acknowledge Grant Nos. NSF PHY-0555325 and NSF-MRSEC DMR-0213745. We
thank Cheng Chin for extremely useful advice, and Yong-Il Shin and W.
Ketterle for helpful discussions.

\bibliographystyle{apsrev}
%\bibliography{Review2}

\end{document}